\documentclass[epj]{svjour_2}
\usepackage{graphicx}

\begin{document}
%\normalsize
%
\title{Study of QGP signatures with the
$\phi \rightarrow {K^+}{K^-}$ signal in Pb-Pb ALICE events}

%\medskip
%\vspace{1.0cm}
%\begin{center}

%\end{center}

%\vspace{1.0cm}
%\begin{center}
%DRAFT \\
%08 Nov 2004
%\end{center}
%\collab{ALICE TOF Group }
\author{A. Akindinov\inst{1},
A. Alici\inst{2,3},
P. Antonioli\inst{3},
S. Arcelli\inst{3,4},
M. Basile\inst{2,3},
G. Cara Romeo\inst{3},
M. Chumakov\inst{1},
L. Cifarelli\inst{2,3,4},
F. Cindolo\inst{3},
A. De Caro\inst{5},
D. De Gruttola{\inst5},
S. De Pasquale\inst{5},
A. Di Bartolomeo\inst{5},
M. Fusco Girard\inst{5},
Yu. Grishuk\inst{1},
C.~Guarnaccia\inst{5},
M. Guida\inst{5},
D. Hatzifotiadou\inst{3},
D.W. Kim\inst{6},
J.S. Kim\inst{7},
S. Kiselev\inst{1},
G. Laurenti\inst{3},
K. Lee\inst{6},
S.C.~Lee\inst{6},
Ye. Lyublev\inst{1},
M.L. Luvisetto\inst{3},
D. Mal'kevich\inst{1},
A. Margotti\inst{3},
A. Martemiyanov\inst{1},
K. Mikhin\inst{1},
R. Nania\inst{3},
F.~Noferini\inst{2,3},
A. Pesci\inst{3},
M. Ryabinin\inst{1},
E. Scapparone\inst{3},
G. Scioli\inst{2,3},
A. Selivanov\inst{1},
S. Sellitto\inst{5},
R. Silvestri\inst{5},
A.~Smirnitskiy\inst{1},
Y. Sun\inst{7},
G. Valenti\inst{3},
I. Vetlitskiy\inst{1},
K. Voloshin\inst{1},
L. Vorobiev\inst{1},
M.C.S. Williams\inst{3},
D. Yakorev\inst{6},
B.~Zagreev\inst{1},
C. Zampolli\inst{2,3},
A. Zichichi\inst{2,3,4}}
\institute{
Institute for Theoretical and Experimental Physics, Moscow, Russia \and
Dipartimento di Fisica dell'Universit\`a, Bologna, Italy \and 
Sezione INFN, Bologna, Italy 
\and Museo Storico della Fisica e Centro Studi e Ricerche ``Enrico
Fermi'', Roma, Italy
\and Dipartimento di Fisica dell'Universit\`a and INFN, Salerno, Italy \and
Department of Physics, Kangnung National University, Kangnung, 
South Korea \and World Laboratory, Lausanne, Switzerland}
%\address[UNI]{}
%\address[BO]{}
%\address[ITEP]{}
%\address[SA]{}
%\address[WLAB]{}
%\address[Korea]{}
%\address[Russia]{}

\vspace{1cm}
\abstract{
The $\phi \rightarrow {K^+}{K^-}$ decay channel
in Pb-Pb collisions at LHC
is studied through a full simulation of the ALICE detector.
The study focuses on possible signatures
in this channel
of quark-gluon plasma (QGP) formation.
On a basis of $10^6$ collisions at high centrality
some proposed QGP signatures are
clearly visible both
in ${K^+}{K^-}$ invariant mass
and transverse
mass distributions.
The high significance of this observation
appears to reside heavily 
on the use of the TOF (Time Of Flight) system of ALICE
in addition to its central tracking detectors. 
\PACS{
      {PACS-key}{discribing text of that key}   \and
      {PACS-key}{discribing text of that key}
     } % end of PACS codes
}
\maketitle
%\vspace{2cm}
\section{Introduction}

High-energy heavy-ion collisions are considered
since a long time (see for example \cite{Jacob})
golden events 
as regards the creation and detection of a quark-gluon 
plasma (QGP).
At SPS a possible evidence for the production
of a new state of matter, 
possibly coinciding with QGP,
has been found \cite{NA,NA2,Ceres,NA57}. 
The results from RHIC heavy-ion experiments
appear to confirm the QGP hypothesis, 
and seem to give evidence of some 
additional property
of the formed plasma \cite{RHICnew}.

The nature and the characteristics of QGP
are in large respects even theoretically unknown. 
This suggests that
a much larger set of measurements
is needed
to recognize what has been really discovered
by the cited experiments and 
to reconstruct the characteristics of QGP.
Presumably only from the convergence
of many different experimental signatures
can the QGP be fully understood and
the QGP interpretation of the data
established 
beyond any doubt. 

One of the expected 
and already detected
signatures is tied to the s-quark 
enhancement
accompanying (partial) chiral symmetry restoration.
This sho\-uld raise the strange and multiply
strange particle content
of the event \cite{Rafelsky&Muller,Rafelsky,Back,Afanasiev_s,AGS,STARratio,PHENIXratio,PHENIXphi}.
 
In this respect the $\phi$ particle is of particular interest because
it is the lightest of the vector mesons with hidden flavour.
The production and decay rates of these particles
are strongly influenced by the OZI (Okubo, Zweig, Iizuka) rule,
that requires the transitions to be described by connected
quark diagrams \cite{Okubo,Lipkin}.
A first consequence is that in case of QGP
a strong enhancement of the $\phi$ rate is expected
as the system has abundance of s-quarks 
in the initial state to feed $\phi$ production.
A second consequence is the small cross section
of the $\phi$ for scattering
with non strange hadrons. This implies that
$\phi$ particles are only slightly affected
by rescatterings with nuclear matter during
the expanding phase, rescatterings that could wash out the
spectral information of the plasma.

In previous works the capabilities of the future ALICE 
experiment at LHC \cite{ALICE}
in extracting the ``normal'' $\phi$ meson signal
were discussed, 
using a full detector Monte Carlo simulation 
\cite{Decaro,DecaroTesi}.
The aim of this paper is to present the results of
a first study, again through full simulation 
of the apparatus,
of the potentialities
of ALICE 
in the detection of QGP signatures
in the $\phi$ signal.

The study of the $\phi$ in the $\phi \rightarrow {K^+}{K^-}$ channel
has some additional problems with respect to
the $\phi \rightarrow {\ell^+}{\ell^-}$ channel
(${\ell}$ stands for leptons).
These are mainly related to the fact that kaons are
subject to strong interactions,
with consequent possible rescatterings with nuclear matter,
and that the $\phi \rightarrow {K^+}{K^-}$ threshold 
(i.e. twice the kaon mass) and
the nominal $\phi$ mass are close to each other, with phase space problems
in the study of possible modifications of the
mass distribution.

On the other hand the $\phi \rightarrow {K^+}{K^-}$ decay
has a branching ratio of 0.49 
whereas the value for any leptonic decay mode is of the order of $10^{-4}$.
In addition kaon identification is performed in ALICE
with low contamination and high efficiency over a large
solid angle within a wide energy range.
These circumstances suggest that at least in the initial stage of
ALICE data taking, the $\phi \rightarrow {K^+}{K^-}$ should be
the most interesting channel for $\phi$ studies.

The present work investigates
the sensitivity of ALICE to QGP signatures connected to the
$\phi$ decay channel into kaons,
in the hypothesis 
that some kind of modifications
are present in the $\phi$ mass distribution 
and/or in the $\phi$ momentum spectrum in the plasma phase
and that they survive, even if largely distorted,
after $\phi$ and kaon propagation. 
While performing this, particular attention is paid
to the gain in the $\phi \rightarrow {K^+}{K^-}$ analysis
coming from the addition to the TPC
(Time Projection Chamber \cite{TPC}) 
of the TOF (Time Of Flight \cite{TOF}) system 
for particle identification (PID).
This in view of the fact that the presence of a powerful
TOF barrel detector,
extending PID
at medium/high momenta \cite{TOF,PID}, will be a key of ALICE.
%(STAR \cite{STARbase} and PHENIX \cite{PHENIXbase}).

\section{Physics hints}
\label{sec2}

An enhancement of s-quark production in connection
with QGP formation has been predicted
since a long time \cite{Rafelsky&Muller,Rafelsky}.
An enhancement of the $\phi/\omega$ ratio, as remarked
in \cite{Shor}, should point to this effect in a
particularly clean way,
but also an enhancement in the $\phi$/hadron
ratio is in general expected.

The s-quark density is readily calculable
in case of non-interacting QGP,
assuming that the equilibrium level 
is reached \cite{Danos,Domokos}.
During the collision however,
even if QGP is formed, ordinary nuclear matter
is also present and,
in any case, QGP evolves to ordinary nuclear matter
at freeze-out.
In addition an s-quark enhancement can be expected also
without QGP formation, as a consequence of pure
nuclear effects, so that
only the details of the enhancement
as a function of nucleon-nucleon center-of-mass energy, \ $\root\of{s_{NN}}$,
could permit to discriminate among QGP and non-QGP scenarios.

It has been suggested that the dependence on \ $\root\of{s_{NN}}$ of
the s-quark content in case of QGP formation could be non
monotonic \cite{Gazdzicky1,Gazdzicky2,Gazdzicky3}, 
this being detectable for example in
the s/$\pi$ ratio.
Experimental confirmation for this prediction
can perhaps be
seen \cite{Afanasiev_non_m,Mischke} 
from the $K/\pi$ measurements 
obtained at AGS/SIS \cite{AGS},
SPS \cite{Afanasiev_non_m} and RHIC 
\cite{STARratio,PHENIXratio,PHENIXphi}.

Then even though the
s-quark enhancement is a very interesting item,
there are some ambiguities
in the evaluation of the expected amount in
case of QGP formation and in the interpretation of the
enhancement already detected so far.
It is generally believed that
this item alone is not 
(and presumably will not be, in the future)
sufficient
to constrain
to QGP interpretation; other items must be
considered in addition.

In the present work we are concerned
with signatures of QGP related to the $\phi$ signal
alone.
This means that we will try to estimate the sensitivity 
with respect to possible changes
in some properties of $\phi$ mesons,
for example in their mass distribution,
assuming a reasonable scenario for s-quark
enhancement at LHC energies.

The $\phi$ signal in the $\phi \rightarrow {K^+}{K^-}$ 
decay channel is extracted from the large
${K^+}{K^-}$ combinatorial background.
Also this background is influenced by
the amount of s-quark enhancement. However
if the $K$ and $\phi$ yields grow in the same
proportion, the $\phi$ signal significance
in first approximation does not change.
A key element in the simulations is therefore
the $\phi/K$ ratio.

According to experimental results 
\cite{STARratio,PHENIXratio,PHENIXphi,Afanasiev_non_m,Mischke},
the\-re is no sign of a significant increase
in the ${K^+}/{\pi^+}$ ratio going from SPS
to RHIC energies, with a kind of plateau already 
starting at the lowest SPS energies.
Different is the situation for the ${K^-}/{\pi^-}$ ratio,
for which a rise from SPS to RHIC energies is visible;
at the highest energies the two ratios, for positively 
and negatively charged particles, are almost equal.

As regards the $\phi$ meson, a monotonic increase
in the $\phi/{h^-}$ ratio with \ $\root\of{s_{NN}}$ 
is present \cite{Mischke}
($h^-$ stands for negative hadrons),
starting from few GeV energies \cite{AGS}
and going through SPS \cite{NA2,Afanasiev_non_m,Mischke} to RHIC
energies \cite{STARphi,Ullrich}.
The $\phi/{K^-}$ ratios found at STAR \cite{Ullrich,STARphi2,Markert}
and PHENIX \cite{PHENIXphi}  
appear however to be roughly
the same as at top SPS energies \cite{NA2}.

The $K/\pi$ ratio within the HIJING generator \cite{HIJING}
is 0.11 at $\root\of{s_{NN}} = 5.5$ TeV,
i.e. a bit lower than the RHIC value ($\sim 0.15$)
\cite{STARratio,PHENIXratio,PHENIXphi}.
The $\phi/{K^-}$ ratio, 
as measured at RHIC
\cite{PHENIXphi,Ullrich,STARphi2,Markert},
is consistent within errors with
the HIJING value of 0.16 at 5.5 TeV.

In our simulations we have used
the HIJING values at $\sqrt{s_{NN}} = 5.5$ TeV 
both for kaon and 
$\phi$ multiplicities.
As we are referring to LHC energies, 
this appears to be a rather conservative assumption.
In fact the quoted experimental results
suggest that above RHIC energies kaon production
presumably saturates, whereas there is 
no
indication in this sense for $\phi$ mesons.
As a consequence
the $\phi/K$ ratio at LHC energies is likely to be 
larger than at RHIC energies.

During heavy-ion
collisions at LHC energies,
$\phi$ generation and decay happen in a heterogeneous
and variable medium
for which two components can be envisaged.
One is purely nuclear, consisting of
nucleons/hadrons at various number densities and
temperatures, but far from $T_c$, the critical temperature
for the onset of QGP transition;
the other one is the foreseen QGP component.
The latter may also include a mixed-phase configuration
(occurring during plasma expansion) in which some
plasma effects still persist.
To the former component belong in particular all 
nucleons/hadrons present
after chemical freeze-out\footnote{When, 
in the evolution of the hadronic
system, inelastic collisions cease and particle ratios
become fixed.}
is reached. Let us call it in the following the
freeze-out component.

To estimate
the relative yield of $\phi$ meson decays
from the two components is a quite complicated
task. It depends on parameters
theoretically largely undetermined
and has no hope of experimental assessment
at present. 
In case of a first order phase transition,
the estimated duration
of the QGP phase is relatively large ($\sim 10$ fm/$c$, comparable
with a vacuum lifetime of $\phi$ mesons of
$\sim 45$ fm/$c$) and it can be expected that
the two yields are comparable \cite{Asakawa}.
This should hold true even if the phase 
transition is instead smooth  
\cite{Asakawa}, since lattice calculations show
that in this case the phase transition is a crossover
very close to a first order one \cite{Brown}.

At LHC energies
the 
$\phi$ mesons if decaying inside the freeze-out 
component
should have 
their free-space
Breit-Wigner mass distribution
centred at
1019.5 MeV/c$^2$ 
with a width 
of 4.3 MeV/c$^2$.
Modifications of the resonance mass,
width and shape can indeed be present in nuclear matter
at high density,
coming from phase space distortions
and dynamical interactions with the medium \cite{Fachini}.
They are expected to be observable, but small.

For decays inside the QGP component
the foreseen situation is very different.
First of all,
as a consequence of partial chiral symmetry
restoration,
the $\phi$ mass distribution may be at sizeably lower
values \cite
{Pisarski,Hatsuda,Asakawa2,Massshift}.
The exact value of the estimated shift
is subject to large uncertainties,
depending on the assumptions made in
the calculations.
The plasma $\phi$ mass could be
10-20\% lower than the nominal mass value,
but also very near to it.

Since the kaon mass is expected to change
very smoothly with temperature \cite{KoSeibert},
the shifted $\phi$ mass can also be well below
the ${K^+}{K^-}$ decay threshold in the plasma.
This implies that in principle the width of
the resonance should decrease in the plasma.
However the width is also influenced by additional 
effects related to the presence of a medium.
These could result from an attractive kaon potential
favouring $\phi$ decays \cite{Shuryak}
(negligible when the $\phi$ mass is well below
the two-kaon mass)
or from $\phi$ interactions with surrounding
hadrons, such as $\pi$, $K$, $\rho$ and $\phi$ itself,
enabling $\phi$ decay modes
otherwise not present in free-space \cite{KoSeibert,Bi}
and changing the $\phi$ spectral function.
As a net result an increase of the width
up to $\sim 10$ MeV \cite{KoSeibert} can fairly be expected.
Further few MeV are expected
to come through the interactions with partons
in the plasma \cite{SeibertKo}.
The calculated width depends on
the effective interaction model chosen to describe
the plasma. Some choices can give even larger widths,
of the order of a few tens of MeV,
as in the case of \cite{Holt}.

On top of this, as we are studying $\phi$ decays
in the $\phi \rightarrow {K^+}{K^-}$ channel,
we must consider
the effects on the observed width 
of kaon collisions with light mesons,
both for the freeze-out and QGP components.
In-medium kaon collisions change
the kaon spectral function
(off-shell kaon final states) 
and, as a consequence,
the $\phi$ decay rate.
In addition,
in-medium 
(QGP or freeze-out)
rescatterings of on-shell kaons
should have an influence on the measured width.
In principle these rescatterings can 
even completely wash out the signal.
In the present work we assume that 
a sizeable
fraction of $\phi \rightarrow {K^+}{K^-}$ decays
occurring in the QGP component
is ``useful'' to produce a signal with an
additional width, 
due
to off-shell kaon effects
coupled with kaon rescattering effects,
comparable
with the original $\phi$ width.
Also in the case of decays in the freeze-out
component, a width larger than the normal one
has been used as a way to describe in-medium effects.

According to theoretical expectations \cite{Asakawa}
as well as to experimental hints \cite{PHENIXphi,STARphi,STARphi2},
the transverse momentum (or transverse mass) distribution of $\phi$ mesons
generated inside the freeze-out component at LHC
should be harder than the 
distribution found at RHIC,
that is with an inverse slope (i.e. temperature) parameter $T$
larger than the RHIC measured value: 
$T = 379 \pm 50 \pm 45$ MeV \cite{STARphi},
or
$T = 363 \pm 8$ (stat.) MeV  
(at 0-5\% centrality, 
with similar values for other centralities) \cite{STARphi2},
or
$T = 366 \pm 11 \pm 18$ MeV (minimum bias) \cite{PHENIXphi}.

On the contrary a much softer distribution
is expected for $\phi$ particles generated
inside the QGP component, with $T \simeq {T_c}$
(independently of the initial temperature \cite{Asakawa})
and typical expected values for ${T_c}$
in the range 160-180 MeV.
Due to the small probability of rescattering
of the $\phi$ mesons after chemical freeze-out \cite{Shor},
the $\phi$ particles generated in the plasma
should retain their energy distribution in the plasma
even when they
subsequently decay in the freeze-out component.
All this suggests that transverse mass\footnote{$m_t = \left (m_{inv}^2 
+ {p_t}^2\right )^{1/2}$.} (${m_t}$) distributions
are another powerful tool, 
in addition to invariant mass ($m_{inv}$) distributions,
to extract evidence for QGP formation from the data
and, in case, to derive $T_c$ from slope measurements 
\cite{Asakawa}.

The expected line-shapes of ${K^+}{K^-}$ invariant and transverse
mass distributions emerging from the described scenario
are then as follows.

The $m_{inv}$ distribution can 
have two peaks, one localized at the normal
(free space) $\phi$ mass value and the other at lower values.
The normal peak
should correspond to $\phi$ particles either
produced and decaying in the freeze-out component 
or produced in the QGP component
but decaying afterwards in the freeze-out one.
The lower mass peak should be formed by
$\phi$ mesons produced and decaying in the plasma.

A ``bridge'' between the two peaks can be expected,
reflecting the possible mixed-phase evolution from plasma
to chemical freeze-out.

The lower mass peak can also
be missing; of course this may occur
when the $\phi$ mass in the plasma is
lower than the two-kaon threshold.
On the other hand, the QGP peak can
also be so near to the normal freeze-out  peak
to be practically undistinguishable.
However,
even in this case, a signature of QGP
formation could be searched for in the
${m_t}$ distribution as we will see later on.
A double component could/should in fact
be present, one (QGP) mostly present at low
${m_t}$ values and the other one (freeze-out)
surely dominant at large values.

\section{Simulation strategy}
\label{sec3}

Orders of ${10^5}$-${10^6}$ events
turned out to be needed to simulate 
a significant signal for the present analysis.
Direct full generation, tracking and reconstruction
of such a huge number of events, 
corresponding to very high multiplicity
final states from central Pb-Pb interactions,
was estimated a rather severe task 
and
simulations have been instead
organized in two steps, as follows.

In the first step,
some hundred central\footnote{The impact parameter range
used in HIJING is 0-3 fm. The resulting
average charged particle density at mid-rapidity
is $\langle dN_{ch}/dy \rangle \sim 6500$.}
HIJING \cite{HIJING} events,
with \ $\root\of{s_{NN}}$ = 5.5 TeV,
were generated,
tracked, digitized and reconstructed, 
making use of detailed
simulations of all ALICE detectors
(see \cite{Decaro,DecaroTesi}). 
In these simulations a magnetic field 
of 0.4 T was used.

From these simulations the relevant parameters
for the subsequent second step
were extracted, as a function of
transverse momentum ($p_t$),
namely
the momentum resolution (in modulus and direction)
and,
both for the TPC
and the TOF,
the PID efficiencies \cite{Decaro,DecaroTesi}.
The momentum is measured by the TPC 
and
the resolutions used are detailed in \cite{Decaro,DecaroTesi}
(for $p_t$ in the range 0.2-9.0 GeV/c, 
the relative $p_t$ 
resolution varies between 1\% and 2\%
and the angular resolutions go from 5.0 to 0.3 mrad
for both polar and azimuthal angles);
the TPC PID capability relies on $dE/dx$
measurements.

In the second step $10^6$ events were generated,
each event containing an average of 850 charged kaons.
This is the number expected from HIJING Pb-Pb central events
at \ $\root\of{s_{NN}}$ = 5.5 TeV in the pseudorapidity ($\eta$)
interval [-1, 1].

The kaons are generated with independent
random choices of rapidity ($y$) and $p_t$.
The $y$ distribution is
a gaussian
with $\sigma =$ 4 \cite{Decaro,DecaroTesi}.
The kaon $p_t$ distribution
is obtained 
through the $m_t$-scaling method
starting from 
the corresponding distribution for pions;
for the latter,
for $p_t < $ 0.5 GeV/c, an $m_t$-scaling 
spectrum has been assumed with $T = $ 160 MeV 
and, for $p_t > $ 0.5 GeV/c, the power law
parametrization reported 
in \cite{CDF} has been chosen.
This gives a kaon
$m_t$ distribution with an inverse
slope parameter varying with $m_t$;
its fitted value being 309 MeV for $\left(m_t-m \right)$ 
in the range 0-1 GeV/c$^2$,
and 390 MeV in the range 1-2 GeV/c$^2$,
where $m$ is the nominal kaon mass.
Both $y$ and $m_t$ distributions  
are the same as used in \cite{Decaro,DecaroTesi}.

In each event an average number of 85 additional 
char\-ged kaons (that is 10\% more)
were generated according to the same distributions  
to take into account
in a simple but conservative manner \cite{TOF}
contamination effects mainly coming from pions.
An average number of 35 $\phi \rightarrow {K^+}{K^-}$
decays per event was also added,
with both kaons
inside the indicated $\eta$ range.
Given the chosen amount of generated kaons,
this number corresponds to the already quoted
$\phi$/${K^-}$ ratio of HIJING. 

%%%%%%%%%%%%%%%%%%%%%%%%%%%%     figure 1  %%%%%%%%%%%%%%%%%%%%%%%%%
%\begin{figure}[ht]
\begin{figure}[h]
\small
\begin{center}
%\resizebox{0.50\textwidth}{!}{%
  \includegraphics[height=7cm,width=9cm]{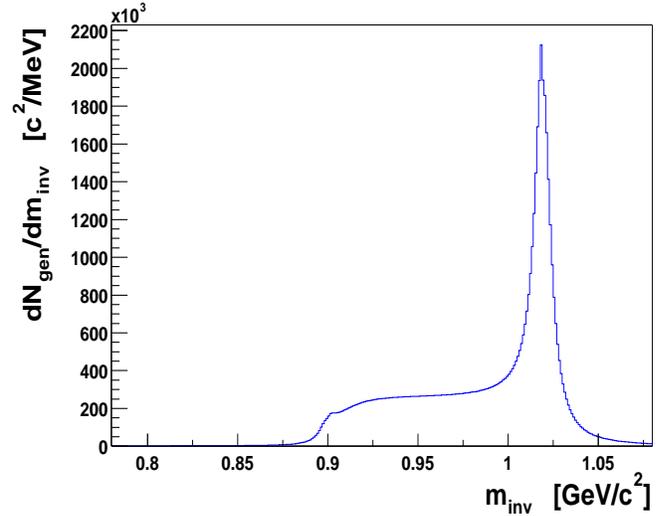}
%}
\caption{\footnotesize 
All-decay $\phi$ mass distribution for case \textit{A} (see text);
the $\phi$ mass inside the plasma is assumed to be
below the two-kaon threshold;
only a small fraction of $\phi$ particles are allowed to decay
inside the plasma fireball; a sizeable fraction is assumed
to decay in a mixed environment 
corresponding to the evolution phase from plasma 
to dense nuclear matter, which gives rise to the so-called 
``bridge'' region.}
   \label{misA}
\end{center}
\end{figure}
%%%%%%%%%%%%%%%%%%%%%%%%%%%%%%%%%%%%%%%%%%%%%%%%%%%%%%%%%%%%%%%%%%%%
%%%%%%%%%%%%%%%%%%%%%%%%%%%%%%%%%   figure2   %%%%%%%%%%%%%%%%%%%%
\begin{figure}[h]
\small
\begin{center}
%\resizebox{0.50\textwidth}{!}{%
  \includegraphics[height=7cm,width=9cm]{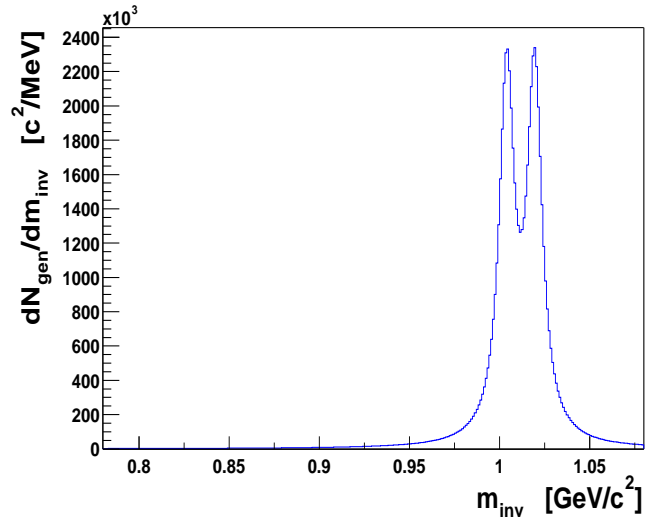}
%}
\caption{\footnotesize 
All-decay $\phi$ mass distribution for case $B$ (see text);
the $\phi$ mass inside the plasma is assumed to be above the 
two-kaon threshold.}
   \label{misB}
\end{center}
\end{figure}
%%%%%%%%%%%%%%%%%%%%%%%%%%%%%%%%%%%%%%%%%%%%%%%%%%%%%%%%%%%%%%%%%%

The $\phi$ particles 
are similarly generated through independent
random choices of $y$ and $p_t$
and they are allowed to decay isotropically
in the center-of-mass frame.
Their $y$ distribution is the same as reported
above for kaons, as in \cite{Decaro,DecaroTesi}.
Their $p_t$ distribution,
differently from \cite{Decaro,DecaroTesi},
was not obtained
through the $m_t$-scaling method used for kaons. 
It required instead a dedicated
approach, according to the purpose of the present work where
we have tried to study
the sensitivity to QGP signatures
when the input $\phi$ mass 
and ${p_t}$ distributions are actually
affected by QGP formation.
To this aim
$\phi$ mass distributions different from
the PYTHIA \cite{PYTHIA} one have been chosen
and
particular ${m_t}$ distributions,
for $\phi$ mesons generated in both the
QGP and the freeze-out components, have been used.

%%%%%%%%%%%%%%%%%%%%%%%%%%%%%%%%%%%%%%    figure3   %%%%%%%%%%%%%%%%%%%
%%%%%%%%%%%%%%%%%%%%%%%%%%%%%%%%%%%%%%    figure3   %%%%%%%%%%%%%%%%%%%
\begin{figure}[ht]
\begin{center}
%\begin{tabular}{cc}
%\hspace{-0.5cm}
%\small
%\resizebox{0.5\textwidth}{!}{%
 \includegraphics[height=9cm,width=9cm]{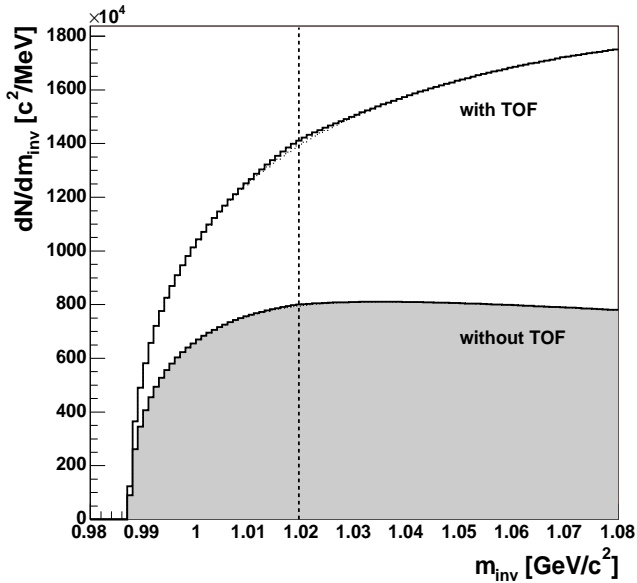} 
%\end{tabular}
\end{center}
   \label{dati}
\caption{\footnotesize
Invariant mass distribution for all opposite-sign
identified kaon pairs in $10^6$ events.
The normalized (see text) background distribution relative
to equal-sign identified kaon
pairs is superimposed.
The results with/without TOF are compared.}
\end{figure}
%%%%%%%%%%%%%%%%%%%%%%%%%%%%%%%%%%%%%%%%%%%%%%%%%%%%%%%%%%%%%%%%%%%%%%%

For each event all kaons were finally
processed according to momentum resolutions
and PID efficiencies extracted from the first step
simulation \cite{Decaro,DecaroTesi}.

Concerning the generation of the $\phi$ invariant mass
spectrum, two very different situations
may appear:
one (call it case $A$) in which
the $\phi$ mass inside the plasma 
(referred to as QGP $\phi$ mass in the following)
is below the two-kaon mass threshold
and another one (case $B$) in which it is above.

For case $A$, the results turn out to be quite insensitive
to the exact value of the QGP $\phi$ mass as well as to the
relative importance between the pure QGP part and 
the freeze-out part
of the $\phi$ mass distribution.
What is crucial instead is the contribution assumed for the ``bridge''
region, corresponding to the
evolution from QGP to chemical freeze-out.

In Fig. \ref{misA} an example of generated $\phi$ mass distribution
representative of case $A$ is reported.
Here all $\phi$ decay modes are included
(with the $\phi$ produced  in the $y$ interval
[-1, 1]) 
and $m_{inv}$ is the invariant mass of any decay products.
Two (QGP and freeze-out) Breit-Wigner peaks,
each with an enlarged width $\Gamma = 10$ MeV/c$^2$, are
assumed. The freeze-out peak is centred at the nominal $\phi$ mass
value ($m_{\mathit 0}$). The additional ``bridge'' region is taken as a uniform (flat)
distribution extending from one peak maximum to the other. 
Normalization is such that the number of
$\phi \rightarrow {K^+}{K^-}$ decays
(strictly speaking, the number of useful $\phi$
decays, in the sense explained in Section \ref{sec2})
is obtained through multiplication
by a factor $\Psi(m)/\Psi({m_{\mathit 0}})$, where
$\Psi(m)$ and $\Psi({m_{\mathit 0}})$ are the
2-body $K^+K^-$ phase space factors, respectively,
at $\phi$ masses $m$
and ${m_{\mathit 0}}$. 
This gives zero 
$\phi \rightarrow {K^+}{K^-}$ actual decays below the two-kaon threshold.
The peak around the chosen QGP $\phi$ mass 
(0.9 GeV/c$^2$)
is, in this example, depressed. 
The population of $\phi$ decays within the ``bridge'' region 
is assumed instead 
equal to that of the freeze-out peak, i.e. the peak 
above the $KK$ threshold, at the nominal $\phi$ mass.
 
For case $B$, the results both in the ${K^+}{K^-}$ invariant mass line-shape
and in the ${K^+}{K^-}$ transverse mass spectrum depend on the
exact location of the chosen QGP $\phi$ mass.

In Fig. 2 
(analogous to Fig. 1)
an example
of $\phi$ mass distribution generated
for case $B$ is reported.
In this example, the QGP $\phi$ mass peak
is at $m = 1004$ MeV/c$^2$, which is approximately
half way between the $KK$ threshold and
the nominal $\phi$ mass.
Here the two Breit-Wigner peaks are assumed
to be equally populated and the additional ``bridge'' region 
strongly depressed.

Turning now to the corresponding ${m_t}$ spectra,  
these are
described
by fixed inverse slope parameters.
For $\phi$ particles generated in the freeze-out
component, the value
$T_{\mathit 0} = 480$ MeV 
(with a spectrum
close to that in \cite{Decaro,DecaroTesi})
has been chosen.
This value is higher than 
the measured RHIC values \cite{PHENIXphi,STARphi,STARphi2}
quoted in Section \ref{sec2}, thus allowing to take into account
the effects of a larger initial temperature
at LHC energies.
The critical temperature for the transition to
QGP has been assumed to be ${T_c} = 180$ MeV,
hence this value 
was chosen for $\phi$ mesons
generated in the plasma \cite{Asakawa}.

For each $\phi$ mass distribution used in the present exercise 
(Fig. \ref{misA} or \ref{misB}), the corresponding
$m_t$ distribution is the sum of only two components.
The first, with inverse slope $T_{\mathit 0}$, corresponds
to $\phi$ meson decays belonging to the freeze-out
Breit-Wigner peak; the second, with inverse slope $T_c$,
to all the other $\phi$ decays belonging to the QGP peak
and the additional ``bridge'' (if any).
As said in Section \ref{sec2}, the 
freeze-out Breit-Wigner peak should in principle collect all $\phi$
decays occurring in the freeze-out, no matter where the $\phi$ mesons
are actually produced (although most of them would likely 
be produced in the freeze-out). Therefore assuming for the whole 
freeze-out peak a unique temperature 
$T_{\mathit 0}$ corresponds to neglect,
as a first approximation, the cases
in which the $\phi$ is generated
in the plasma (or in the ``bridge'') and decays
in the freeze-out.

\section{Results and discussion}

Figure \ref{dati} shows the inclusive $m_{inv}$ distribution for all identified
${K^+}{K^-}$ pairs relative to the $10^6$ simulated events
obtained as described in Section \ref{sec3}, using as input the 
single-peak $\phi$ mass spectrum of Fig. \ref{misA} (case $A$). 
The ALICE
detector effects are taken into account and 
two configurations are compared: one in which the TOF system is not present
in the detection and PID chain (``without TOF''),
the other in which the TOF system is also considered
(``with TOF'').
The difference in the ${K^+}{K^-}$ pair population 
for the two cases is evident; 
there is roughly a ratio 1.8 in the regions just
below and above 
the nominal
$\phi$ mass.
This difference is obviously
determined by 
the enlargement, thanks to the TOF, of
the momentum region
where kaons can be identified.

A significant $\phi$ signal can be extracted after 
background subtraction.
For the background
we have used analogous $m_{inv}$ distributions corresponding to
all equal-sign identified kaon pairs\footnote{The 
like-sign $({K^+}{K^+}$, ${K^-}{K^-})$ 
background is normalized to the ${K^+}{K^-}$ spectrum 
in the mass regions below
and above the $\phi$ peak.}. The result is shown in Fig. 4.
%%%%%%%%%%%%%%%%%%%%%%%%%%%%%%%%%%%%%%%%%%%%%%  figure4  %%%%%%%%%%%%%%%%
\begin{figure}[h]
\small
\begin{center}
%\resizebox{0.50\textwidth}{!}{%
  \includegraphics[height=7cm,width=9cm]{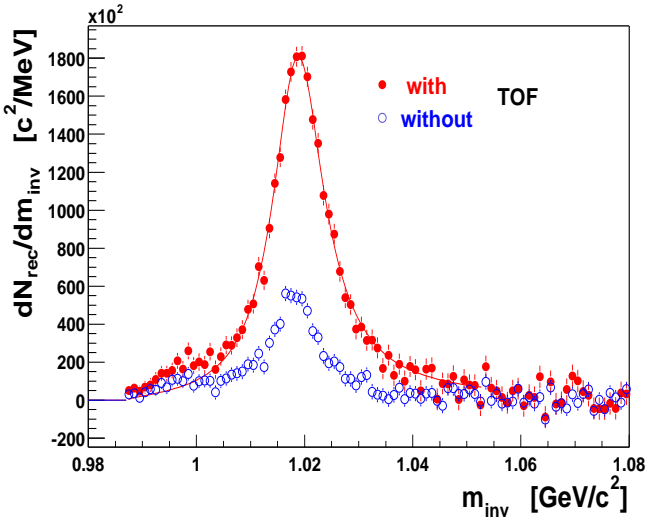}
%}
\caption{\footnotesize
Extracted signal after background subtraction.
The assumed $\phi$ mass input spectrum is that of Fig. 1
(when the $\phi$ mass inside the plasma is below the two-kaon threshold).
The results with/without TOF are compared. 
}
   \label{segnaleA}
\end{center}
\end{figure}
%%%%%%%%%%%%%%%%%%%%%%%%%%%%%%%%%%%%%%%%%%%%%%%%%%%%%%%%%%%%%%%%%%%%%%%%
This figure actually represents what can happen
when the QGP $\phi$ mass is below the $KK$ threshold.

In Fig. \ref{segnaleA}, again a comparison is made
between the results with and without the TOF system.
The ratio between the two peak heights is about 3.3,
that is significantly larger than the above 1.8 ratio
between the two background levels
(with and without TOF, in Fig. \ref{dati}); 
the $\phi$ signal significance
is correspondingly enhanced by a factor
$\sim 3.3/\sqrt{1.8} \sim 2.5$.
This shows that the larger momentum acceptance
for PID, only achievable when the TOF is included in the analysis, 
reflects itself in a larger
signal/background ratio, due to the different spectral
properties of background kaons 
with respect to kaons coming from $\phi$ meson decays.

The fit to the data (with TOF) 
reported in Fig. \ref{segnaleA}
is a Breit-Wigner
weighted with the appropriate phase space factor.
The fit parameters, $m_{\mathit 0} =1018.6$ MeV/c$^2$
and $\Gamma = 11.5$ MeV/c$^2$, are in accordance with
the input parameters used in this exercise and with the 
excellent mass resolution of the detector \cite{Decaro,DecaroTesi}.
A small excess of events with respect
to the fit is visible, 
on the left side of the peak;
this excess comes from 
$\phi \rightarrow {K^+}{K^-}$ decays
in the ``bridge'' region defined in Sections \ref{sec2} and 
\ref{sec3}.

\begin{figure}[h]
\small
\begin{center}
%\resizebox{0.50\textwidth}{!}{%
  \includegraphics[height=7cm,width=9cm]{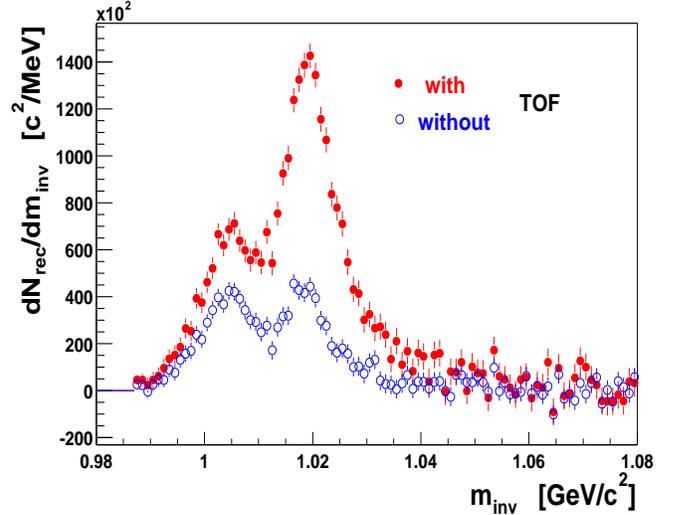}
%}
\caption{\footnotesize
Same as Fig. \ref{segnaleA} but relative to the $\phi$ mass
input spectrum of Fig. 2 (when the $\phi$ mass
inside the plasma is  
above the two-kaon threshold).
}
   \label{segnaleB}
\end{center}
\end{figure}

Figure \ref{segnaleB} shows the ${K^+}{K^-}$ invariant mass distribution
for the signal after background subtraction,
when the $\phi$ mass inside the plasma
is above the $KK$ threshold (case $B$ of Section \ref{sec3});
this figure corresponds now to the $\phi$ mass input spectrum
of Fig. 2. 
%%%%%%%%%%%%%%%%%%%%%%%%%%%%%%%%%%%%%%    figure5    %%%%%%%%%%%%%%%%%%%%%%%
%%%%%%%%%%%%%%%%%%%%%%%%%%%%%%%%%%%%%%%%%%%%%%%%%%%%%%%%%%%%%%%%%%%%%%%%%%%%
Here a double peak structure appears.
One can notice that, with TOF,
the QGP $\phi$ mass peak is depressed with respect
to the freeze-out peak,
despite the fact that
the two peaks are equal
in the input spectrum of Fig. \ref{misB},
while, without TOF, the two peaks remain balanced.
This is a consequence
of the different phase space factors for
the two peaks, combined with the different PID efficiencies vs.
momentum for the TPC and TOF detectors.

When the peaks are too near,
they become practically undistinguishable
and other ways should be searched for
in order to disentangle a QGP contribution.
Table \ref{s_b_lpt} reports the $\phi$ signal significance (within
$\pm \Gamma/2$ from the mean, with $\Gamma = 10$ MeV/c$^2$) 
for both peaks and 
for various choices of the QGP $\phi$ mass. The results are given 
with and without TOF\footnote{It can be noticed that the with/without
TOF significance ratios in this table $-$ from top to bottom $-$
increase from 1.2 to 1.5 for the QGP peak and decrease from 2.7 to 
1.9 for the freeze-out peak. Again this is a consequence of the 
different temperatures assumed for the two (QGP and freeze-out) components
and of the different performances vs. $p_t$ of the two (TPC and TOF) 
detectors. In particular this explains why when the QGP and freeze-out
peaks are superimposed (last table row) 
the significance gain due to the TOF is lower
than for the single freeze-out peak of Fig. \ref{segnaleA}.}.

%%%%%%%%%%%%%%%%%%%%%%%%%%%%%%%%%%%%%%%%%%%%%%%%%%%%%%%%%%%%% Table %%%%%%%%%%%%%%%%%%%%%%%%%%
\begin{table}[ht]
  \caption{$\phi$ signal significance for case $B$ (see text) in 
$10^6$ events; the $\phi$ mass inside the plasma is assumed to be above 
the two-kaon threshold and the all-decay $\phi$ mass input spectrum 
to contain equally populated QGP and freeze-out peaks.}

  \vglue.3cm
  \label{s_b_lpt}

  \begin{center}
  {\renewcommand{\arraystretch}{1.}
    \begin{tabular}{|c|c|c|c|c|}      \hline
      QGP $\phi$ mass & \multicolumn{2}{c|}{significance}  & \multicolumn{2}{c|}{significance}  \\ 
        \ (GeV/c$^2$) & \multicolumn{2}{c|}{without TOF} & \multicolumn{2}{c|}{with TOF} \\
        \cline{2-5}\             & QGP  &  freeze-out  &  QGP   &  freeze-out\\ \hline\hline
      0.99            & 32.8 & 41.3         & 39.2   & 112.2  \\ \hline
      1.              & 41.9 & 37.4         & 53.0   & 98.7   \\ \hline
      1.01            & 48.7 & 43.3         & 74.0   & 101.5  \\ \hline
      1.02 (nominal)  & \multicolumn{2}{c|}{67.7}        &  \multicolumn{2}{c|}{128.1}\\ \hline
    \end{tabular}}
  \end{center}
\end{table}
%%%%%%%%%%%%%%%%%%%%%%%%%%%%%%%%%%%%%%%%%%%%%%%%%%%%%%%%%%%%%%%%%%%%%%%%%%%%%%%%%%%%%%%%%%%%%%
One can see from this table that the signal significance
on a basis of $10^6$ collision events
is actually very high.
In principle, 
with a statistics of 
only some 20,000 events,
a significant $\phi$ signal should already appear.
The point is however that 
important handles
to disentangle a QGP signal  
are the details of the
invariant mass line-shape and, as we will see next,
of the transverse mass distribution.
In view of this a superabundant significance 
of the signal is mandatory. 

\begin{figure}[h]
\small
\begin{center}
%\resizebox{0.50\textwidth}{!}{%
  \includegraphics[height=7cm,width=9cm]{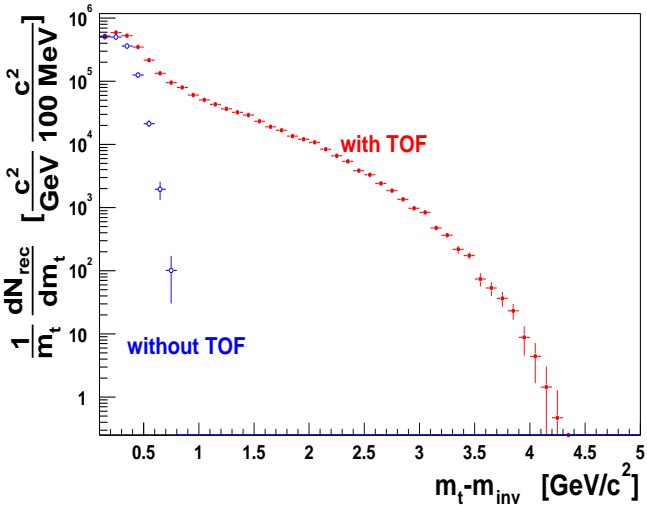}
%}
\caption{\footnotesize
$(m_t-m_{inv})$ distribution of the signal after background subtraction.
The assumed $\phi$ mass input spectrum is that of Fig. \ref{misA},
corresponding to a plasma $\phi$ mass below the two-kaon threshold. The results
with and without TOF are compared.
}
   \label{tmass1}
\end{center}
\end{figure}

As said,
a promising signature of QGP formation
could show up in ${m_t}$ distributions.
Figure \ref{tmass1} reports the ${K^+}{K^-}$ 
($m_t - m_{inv}$) distributions relative to the signal in 
Fig. \ref{segnaleA}
(referring to the input spectrum of Fig. 1)
for the cases with and without TOF.
%%%%%%%%%%%%%%%%%%%%%%%%%%%%%%%%%  figure6   %%%%%%%%%%%%%%%%%%%%%%
%%%%%%%%%%%%%%%%%%%%%%%%%%%%%%%%%%%%%%%%%%%%%%%%%%%%%%%%%%%%%%%%%%%
The background is subtracted using the same
procedure as for Fig. \ref{segnaleA}. The distributions are not corrected for
detector ${K^+}{K^-}$ transverse mass acceptance\footnote{This to allow 
the cleanest possible comparison
(i.e. without introducing additional biases deriving
from the fact that the correction for $m_t$ acceptance
maintains a dependence from the input spectra
assumed for $\phi$ production).},
just to show how large is the improvement 
obtained by adding the TOF. In particular, the region
with $(m_t-m_{inv}) >$ 0.8 GeV/c$^2$ cannot be explored
without the inclusion of this detector \cite{DecaroTesi}.

In Fig. \ref{tmass1} one can see that
the $m_t$ acceptance in the configuration without TOF
falls off very rapidly for $(m_t-m_{inv})$ below 1 GeV/c$^2$.
In these circumstances uncertainties in $m_t$ acceptance
would heavily interfere with genuine $m_t$ spectra
and make the QGP discrimination a very difficult task,
contrary to what happens
if the PID capability of the TOF is enabled.

\begin{figure}[h]
\small
\begin{center}
%\resizebox{0.50\textwidth}{!}{%
  \includegraphics[height=7cm,width=9cm]{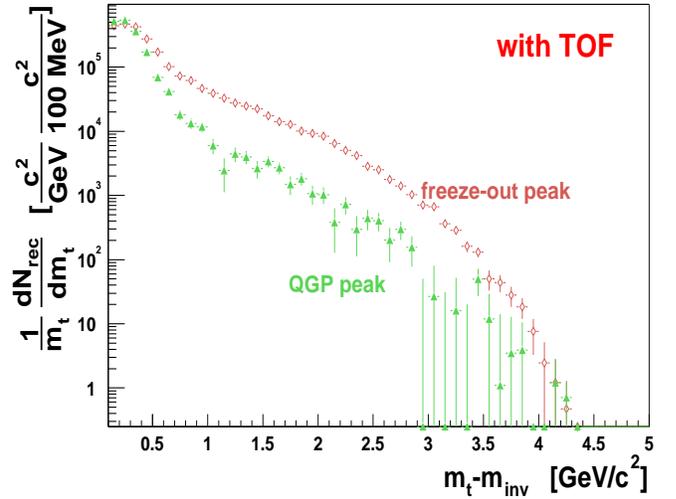}
%}
\caption{\footnotesize
$(m_t-m_{inv})$ distribution of the signal after background subtraction,
in two
different regions of invariant mass (see text and Fig. \ref{segnaleB}): 
one around the vacuum
value (freeze-out peak) and the other one
around the plasma value (QGP peak).
The assumed $\phi$ mass input spectrum is that of Fig. \ref{misB},
corresponding to a plasma $\phi$ mass above
the two-kaon threshold.
}
   \label{tmass2}
\end{center}
\end{figure}

Figure \ref{tmass2} allows to determine the $m_t$ region
relevant for QGP discrimination.
%%%%%%%%%%%%%%%%%%%%%%%%%%%%%%%%  figure7   %%%%%%%%%%%%%%%%%%%%%%%%%
%%%%%%%%%%%%%%%%%%%%%%%%%%%%%%%%%%%%%%%%%%%%%%%%%%%%%%%%%%%%%%%%%%%
The figure refers to the input spectrum of Fig. \ref{misB}
and shows the ${K^+}{K^-}$ $(m_t-m_{inv})$ distributions
(with TOF)
for the signal in Fig. \ref{segnaleB},
after background subtraction, in two $m_{inv}$ regions:
one corresponding to the QGP and the other
to the freeze-out Breit-Wigner peak. Each region 
in Fig. \ref{segnaleB} is selected within $\pm\Gamma/2$
around the corresponding $m_{inv}$ peak value. 
Here again the distributions
are not corrected for detector ${K^+}{K^-}$ transverse mass acceptance.
The difference in slope of 
the two distributions
is clearly evident up to 
$({m_t}-m_{inv}) \simeq$ 1.5 GeV/c$^2$,
with a softer spectrum relative to the QGP peak.
Above this value the shapes of the distributions become quite similar.
This indicates that the contamination from
freeze-out decays within the QGP peak range
grows with $(m_t-m_{inv})$ and becomes nearly 100\%
above that value.

The input spectrum of Fig. 2 has been chosen
as an example of a particularly clean situation,
to show the principle of the discrimination
between
QGP effects and freeze-out effects. 
Nevertheless, using the present $10^6$ event data sample
and taking advantage of the TOF PID, it can be shown
that 
a similar discrimination 
in $m_t$ spectra
can be achieved for every location
of the QGP peak above the $KK$ threshold (assuming
for definiteness equal heights for the two peaks
in the all-decay $m_{inv}$ distribution).
When the two peaks are very near or exactly superimposed,
the two $m_t$ spectra refer 
to the same $m_{inv}$ regions around the nominal $\phi$ mass value;
in this case QGP effects are still visible as a change in the slope
of the spectrum, becoming harder 
above $({m_t}-m_{inv}) \simeq$ 0.5 GeV/c$^2$.

Such an $m_t$ discrimination thanks to the TOF is in principle 
still viable when the QGP $\phi$ mass
is below the $KK$ threshold, as in the situation of
Fig.s 1 and 4, 
even though with smaller statistical
significance.
The two $m_{inv}$ regions to be compared in this case
are the freeze-out peak and a narrow region just above the
$KK$ threshold.

As also ordinary nuclear matter can give some shift
in the resonance mass as well as a distortion in its
shape \cite{Fachini}, the presence of anomalies
in invariant mass distributions of the kind described
in Figs. \ref{segnaleA} and \ref{segnaleB} could fail to
be a compelling indication
of QGP formation.
Then the use of the additional $m_t$ information 
turns out
crucial to obtain such an indication,
whichever the invariant mass line-shape is.
 
In the results shown above the signal has been extracted,
as we have seen,
through a procedure of uncorrelated background subtraction.
Background is however expected
to be uncorrelated only at first approximation.
As there is no knowledge of particles
decaying to ${K^+}{K^-}$ with mass
near the nominal $\phi$ mass,
the most interesting sources of correlated
background potentially relevant for
the present analysis are particles with
$\pi p$, $\pi K$ or $\pi\pi$ decays where one or
both decay products are misidentified as kaons.

The probability that the PID procedure
misidentifies a pion or a proton as a kaon 
does not exceed the few \% level \cite{Decaro,DecaroTesi}. 
This rules out any significant
contribution from $\pi \pi$ or $\pi p$ 
decays of abundantly produced nearby particles 
such as $K_S^0$ or $\Lambda^0$ 
(moreover, for the former, kinematics pushes this contribution 
mainly above the $m_{inv}$ region of interest).
Hence in practice
only $\pi K$ decays can have a relevance.
For this kind of decays the best 
candidate is $K^{0*}$. 
But again, in addition to a $\sim 10^{-2}$ reduction
factor from PID efficiency, almost all
contamination decays from this meson
are
shifted to $m_{inv}$ values larger than the nominal $\phi$ mass
due to kinematics.
The assumption of uncorrelated background
appears then fairly safe. 

A last comment regards
final-state kaon interactions.
All the results obtained reside on the assumption
that a non negligible number of $\phi$ decays  
have the kaons coming out
from the plasma with momenta resembling
their initial momenta (the
so called useful $\phi$ decays, see Section \ref{sec2}).  
To change this number, i.e. to change the relative importance
of the QGP and freeze-out components in the
all-decay input spectra (Figs. \ref{misA} and \ref{misB}),
implies to change the significance of
QGP discrimination.
With a reduced number of useful decays
this discrimination can of course
become impossibile.

On the other side it seems quite difficult
to reproduce the described signatures
of QGP formation through 
re-interactions of kaons 
(or of pions and protons misidentified as kaons)
coming from
decays of resonances other than the $\phi$.
Re-interactions will presumably spread kaon (and other particle) 
momenta and distort their spectra, but it seems
unlikely that they would be able to mimic combined signatures 
of the kind described herein 
in both the invariant mass and transverse mass
distributions.
From this one can infer that
if such signatures 
are observed
they would strongly point to an origin not easily reducible 
to final-state
interaction effects, and contribute 
to compel to a QGP scenario. 

\section{Conclusions}

In this work the sensitivity
of ALICE
to QGP signatures in Pb-Pb collisions has been investigated
through full detector simulation
by studying the $\phi \rightarrow {K^+}{K^-}$
decay channel.

The impact of QGP formation
on two quantities, the mass and the
transverse momentum of $\phi$ mesons produced
inside the plasma, has been considered. 

On a basis of $10^6$ central events,
a discrimination 
of QGP effects has been shown to be possible
for various choices of $\phi$ mass inside the plasma,
even below
the $KK$ threshold.
This can be achieved by looking 
at secondary peaks or distortions
of the invariant mass distribution
of the extracted signal
in the region below the nominal $\phi$ mass
and at the spectral properties 
(transverse mass distribution)
of this signal.

The combined study of both the $m_{inv}$ and the $m_t$ distributions
of $\phi$ mesons has been identified as a powerful tool of investigation.

The significance of the results
depends on various theoretical unknowns such as
the QGP $\phi$ mass value, the duration
of the QGP phase, the evolution mechanism of the system
from plasma to chemical freeze-out,
the particle ratios and spectra,
among others.
 
Experimentally this significance
would strongly benefit from the TOF system
for particle identification.
The overall $\phi$ signal significance 
is actually found to be enhanced by a factor 2-3 thanks to the TOF;
moreover, 
this gain in significance strongly increases 
with $m_t$ while,
in the configuration without the TOF,
no sensitivity is
present
when $(m_t-m_{inv})$ is above
0.8 GeV/c$^2$.
This means that the effectiveness
in QGP discrimination
in terms of 
$m_t$
appears to depend crucially
on the presence of the TOF PID capability.

\end{document}